\documentclass[aps,prl,twocolumn,showpacs,superscriptaddress]{revtex4-1}
\usepackage{amsmath}
\usepackage{amsfonts}
\usepackage{amssymb}
\usepackage{bm}
\usepackage{color}
\usepackage{graphicx}
\newcommand{\be}{\begin{eqnarray}}
\newcommand{\en}{\end{eqnarray}}
\newcommand{\bes}{\begin{subequations}}
\newcommand{\ens}{\end{subequations}}
\newcommand{\ben}{\begin{eqnarray*}}
\newcommand{\enn}{\end{eqnarray*}}
\newcommand{\beq}{\begin{equation}}
\newcommand{\enq}{\end{equation}}

\newcommand{\f}{\frac}

\newcommand{\bi}{\begin{itemize}}
\newcommand{\ei}{\end{itemize}}
\renewcommand{\v}{\varepsilon}

\begin{document}
\title{Oscillators with Parametrically Excited Nonlinearity: Resonance, Anti-resonance and Switch}
\author{Sagar Chakraborty}
\email{sagar@nbi.dk}
\address{NBIA, Niels Bohr Institute, Blegdamsvej 17, 2100 Copenhagen \O, Denmark}
\author{Amartya Sarkar}
\email{amarta345@bose.res.in}
\address{Department of Theoretical Sciences, S. N. Bose National Centre for Basic Sciences, Salt lake, Kolkata 700098, India}
\begin{abstract}
We discover presence of a hitherto unexplored type of resonance in a parametrically excited Van der Pol oscillator.  The oscillator also possesses a state of anti-resonance. In the weak non-linear limit, we explain how to practically get a complete picture of different states of limiting oscillations present in the oscillator when the non-linear term therein is excited by an arbitrary $2\pi$ periodic function of time. We also illustrate how two such oscillators can be coupled to behave like a two-state switch allowing an sharp change of value of amplitude for stable oscillations from one constant to another.
\end{abstract}
\maketitle
Oscillators --- linear and non-linear --- are ubiquitous in mathematical models in almost any research area and, thus, resonance is one of the most common phenomena around us. A simple harmonic oscillator (or \textit{any} oscillator, for that matter) or a system of such oscillators is said to be at resonance when it oscillates with far greater amplitude at certain frequencies (resonant frequencies) than at others in the presence of an external driving periodic force which not necessarily be large. Even in absence of any driving force, the simple harmonic oscillator can have exponentially growing oscillations under certain conditions, for instance, when it is parametrically excited meaning that its parameter oscillates with time. Such parametric resonance \cite{NM} is well studied in the context of Mathieu equation appearing across disciplines over the last century. Also of interest and far more generality and complexity are the nonlinear parametric equations of the form \cite{V}:
\be
\ddot{x}+k\dot{x}+[\omega_0^2+p(t)]\left(x+\sum_{i=1}^N\alpha_i x^{1+i}\right)=0\,.
\en
Here, dot denotes derivative w.r.t. time, $\omega_0$ is the natural frequency of the oscillator, $p(t)$ is a periodic function and $k$ the damping coefficient. Damped Hill equation is recovered when $\alpha_i=0$. While for $p(t)\propto\cos(2\omega_0t)$ in Hill equation, we have the well-known (linearly damped) Mathieu equation. These parametric excitations, when favourable, set up \textit{instabilities} in the center-type oscillations (i.e., amplitude of the unforced oscillation is different for different initial conditions).

For exploring the effect of parametric excitation on self-sustained limit-cycle-type oscillations, Li\'enard equation given by
\be
\ddot{x}+f(x)\dot{x}+g(x)=0,
\en
where $f(x)$ and $g(x)$ are continuously differentiable functions on real line, is a natural candidate for investigation. This is because under suitable conditions, Li\'enard theorem (see e.g. \cite{P01}) guarantees a stable and unique limit cycle surrounding the origin in Li\'enard systems. One of the simplest and very frequently studied Li\'enard system is the Van der Pol oscillator [Eq. (\ref{penvofull}) with $p(t)=1$]. Van der Pol equation, along with its modified forms, have proved to be of practical importance in various physical, chemical and biological sciences \cite{F61,KS98, RS93} as well as in economics \cite{AK86}, and of late, even in seismology \cite{CEGP99}. Parametric excitation on Van der Pol oscillator has been researched in the context of Mathieu-Van-der-Pol equation \cite{VM1,VM2,VM3,VM4}, however the parameter in front of the non-linear term in the equation remains constant therein. This constraint, as we shall see below, when relaxed opens up a plethora of hitherto unexplored exciting possibilities such as an unconventional resonance and anti-resonance phenomena. In passing, it may be mentioned that anti-resonance \cite{E84}, occurring in multi-degree-of-freedom systems at frequencies in between two resonating frequencies, is an important phenomenon studied in the context of seismic protection of structures. Also, among many other applications of anti-resonance, driving piezoelectric motor at anti-resonant frequencies has practical advantages \cite{U98}. 
%

Here, we embark on a study of an equation for the ``{\bf p}arametrically {\bf e}xcited {\bf n}onlinearity in {\bf V}an der Pol {\bf o}csillator" (``\textbf{penvo}" hereafter) defined as
\be
\ddot{x}+\v p(t)(x^2-1)\dot{x}+\omega_0^2x=0\,,\label{penvofull}
\en
where $p(t)$ is a $2\pi$-periodic function and, hence, possesses following Fourier series:
\be
p(t)=\f{a_0}{2}+\sum_{n=1}^{\infty}a_n\cos(nt)+\sum_{n=1}^{\infty}b_n\sin{(nt)}\,.\label{fourier}
\en
Very interesting results, that form the core of the study presented here, and illuminating insights can be gained by using a rather simplified truncation of the series (\ref{fourier}): $p(t)=a_0/2+a_n\cos(nt)$. Below we write down the equation for the \textit{simplified penvo} thus obtained:
\be
\ddot{x}+\v [1+\gamma\cos(\Omega t)](x^2-1)\dot{x}+x=0\,.\label{penvo}
\en
Here, $\gamma$ is a real number. We have rescaled time $t$ and parameter $\v$ in order to fix $\omega_0=1$ and $a_0=2$ without any loss of generality. Also, we have generalised integer $n$ to a positive real number $\Omega$ in order to increase the scope of our study. We shall come back to penvo [Eq. (\ref{penvofull})] again at a later stage. We remark here that we shall solely work in the weak non-linear limit: $|\v|\ll1$ and positive $\v$ where the limit cycle in Van der Pol equation [Eq. (\ref{penvo}) with $\gamma=0$] is stable. Unless otherwise specified, $\v$ has been fixed at $0.001$.
\begin{figure}
\includegraphics[width=4.2cm]{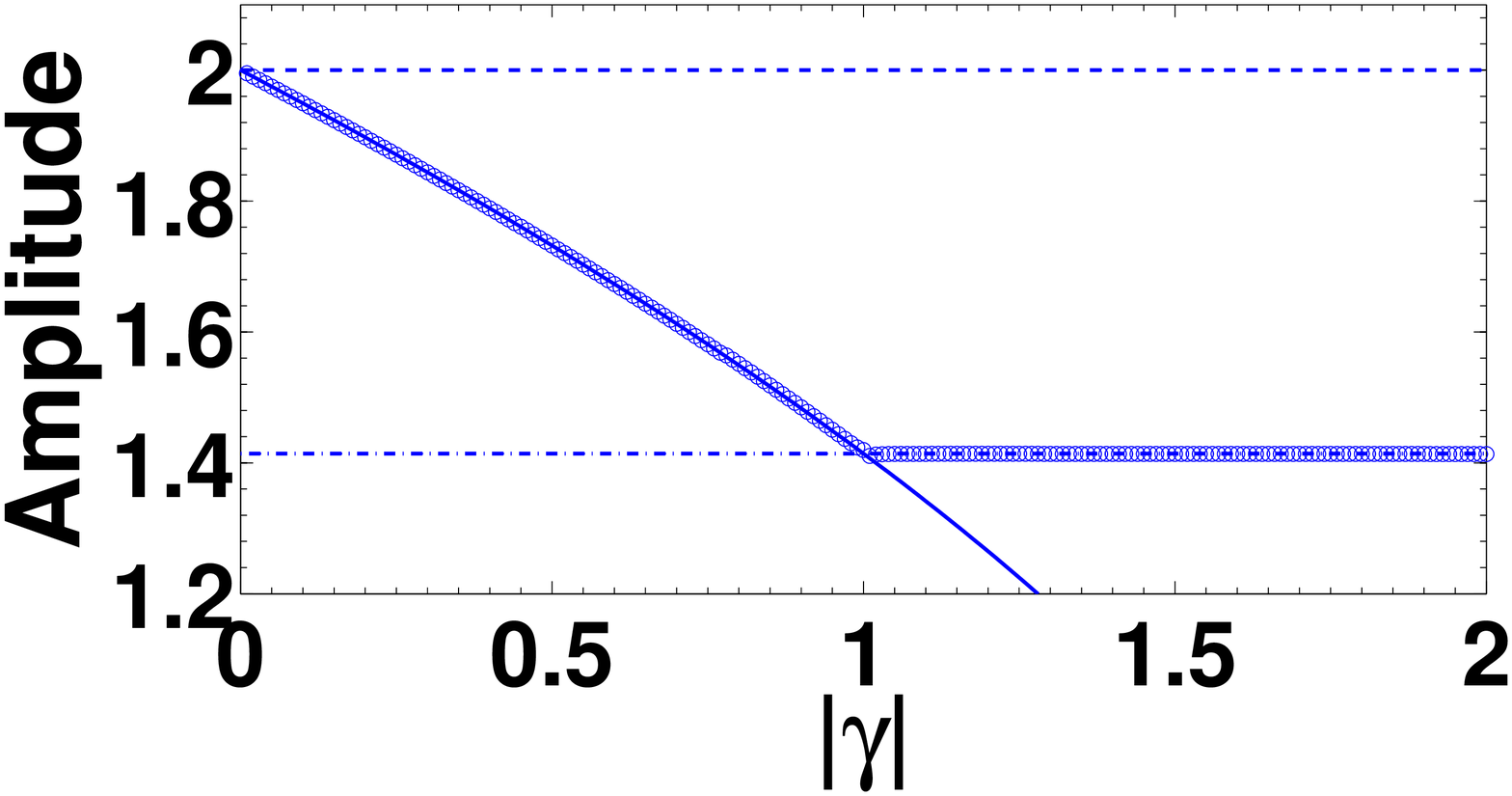}
\includegraphics[width=4.2cm]{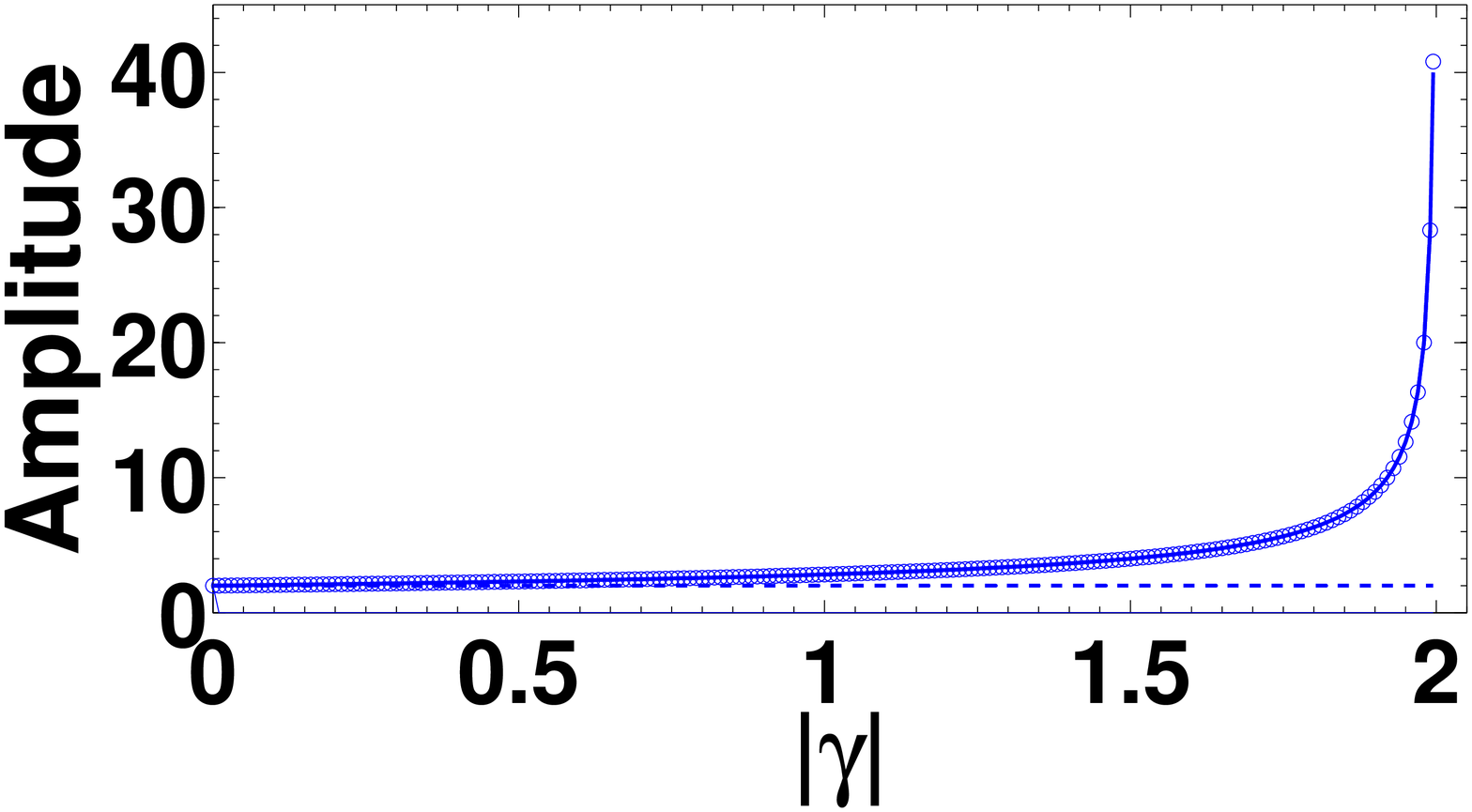}
\caption{Anti-resonance and resonance in simplified penvo: $\Omega=2$ (\textit{left}) and $\Omega=4$ (\textit{right}). The solid line in the left figure is given by $\sqrt{(4-2|\gamma|)}$  and by $\sqrt{8/(2-|\gamma|)}$ in the right figure. The dot-dashed and the dashed lines have the constant values of $\sqrt{2}$ and $2$ respectively. Circular markers which collectively appear as dense thick lines are the numerical data points in excellent fit with theoretical predictions.}\label{f:Fig.1}
\end{figure}

We begin with briefly introducing the main numerical results that are typical of the simplified penvo. For any $\Omega\ne2\textrm { or }4$, double precision numerics show that Eq. (\ref{penvo}) allows for a solution corresponding to a stable limit cycle of radius $2$ --- accurate upto 5\% of relative deviation. When $\Omega$ is irrational, as one may guess, quasiperiodicity can also be seen in the stable oscillations. $\Omega=2$ gives rise to oscillations of amplitude \textit{less than} $2$ and decreasing with increasing $|\gamma|$ until $|\gamma|$ is unity, after which amplitude saturates at a value $\sqrt{2}(<2)$. This can be, aptly, considered as an anti-resonance phenomenon.  At $\Omega=4$, the simplified penvo shows resonance: as $|\gamma|$ is increased from $0$ to $2$ the oscillations' amplitude increase monotonically and always remain \textit{greater than} $2$. Readers are referred to Fig. \ref{f:Fig.1}.

It is worth noting that while to see the resonance one doesn't need an external driving force, this is also unlike the parametric resonance seen in the Mathieu oscillator in at least following two respects: (i) the resonance in simplified penvo is \textit{not} an instability phenomenon where amplitude of the oscillations increases with \textit{time}, and (ii) for resonance the parametric periodic forcing has a period \textit{one-quarter} of the natural period of the simplified penvo where the angular frequency of the limiting oscillations for all $\Omega$ is $1$ up to the first subleading order in $\v$.

In order to explain the numerical results, we resort to the harmonic balance method \cite{JS}. We assume a highly truncated Fourier series expansion of the periodic solutions under question:
\be
x(t)=a(t)\cos(t)+b(t)\sin(t)\,,\label{tfs}
\en
where we assume slow enough time dependence of coefficients $a$ and $b$ so that we can ignore $\ddot{a},\ddot{b},\varepsilon\dot{a}$ and $\varepsilon\dot{b}$ ($|\varepsilon|<<1$) w.r.t. $\dot{a}$ and $\dot{b}$. After substituting Eq. (\ref{tfs}) in the simplified penvo [Eq. (\ref{penvo})], one equates the coefficients of $\cos(t)$ and $\sin(t)$ to zero. Consequently, we can find the flow equations for $a$ and $b$. The trajectories of the flow equations are, thus, easily visualized in the Van der Pol plane i.e. the phase plane for $a$ and $b$. It is convenient to consider three distinct cases.
\begin{figure}
\includegraphics[width=4.2cm]{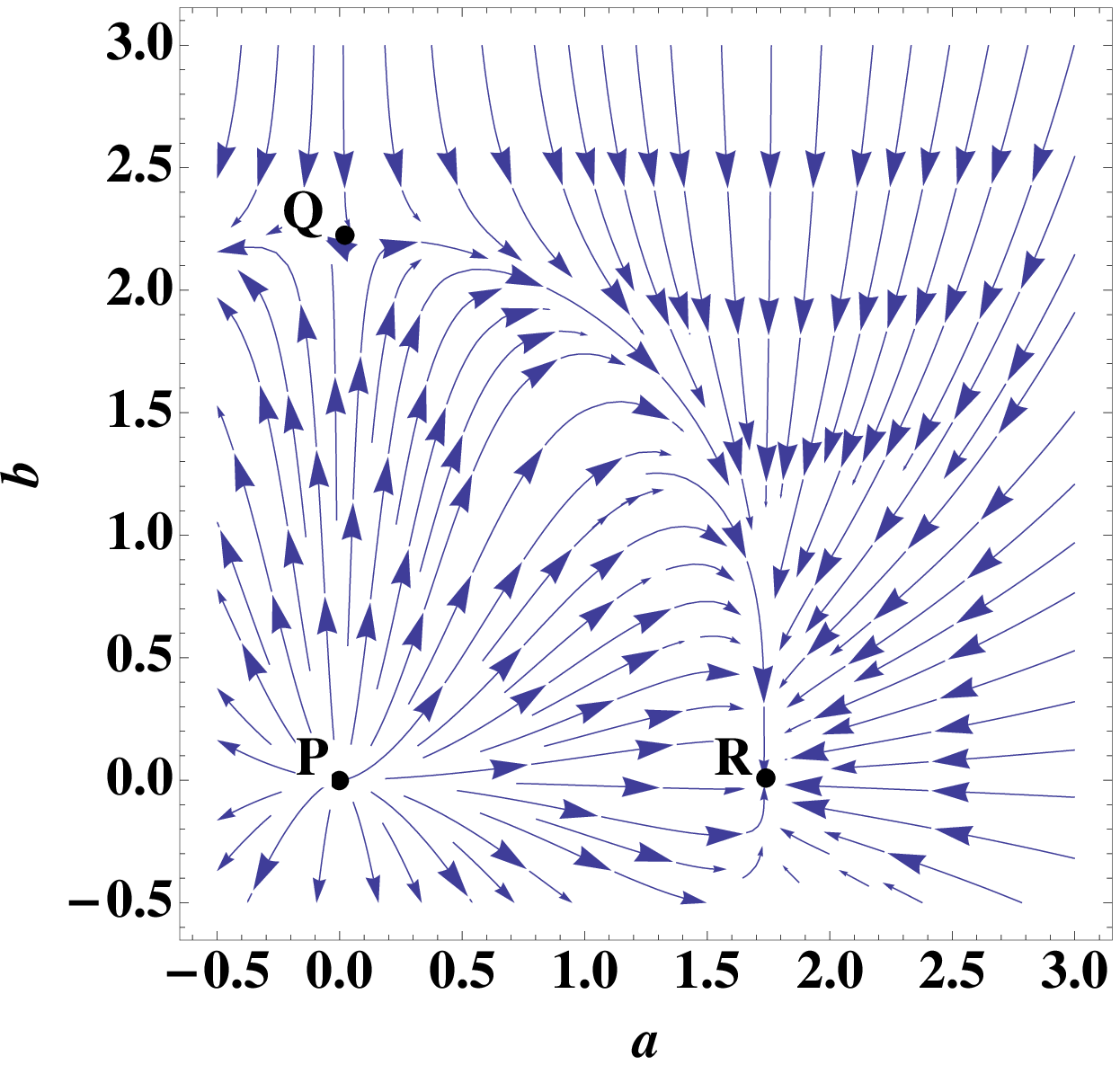}
\includegraphics[width=4.2cm]{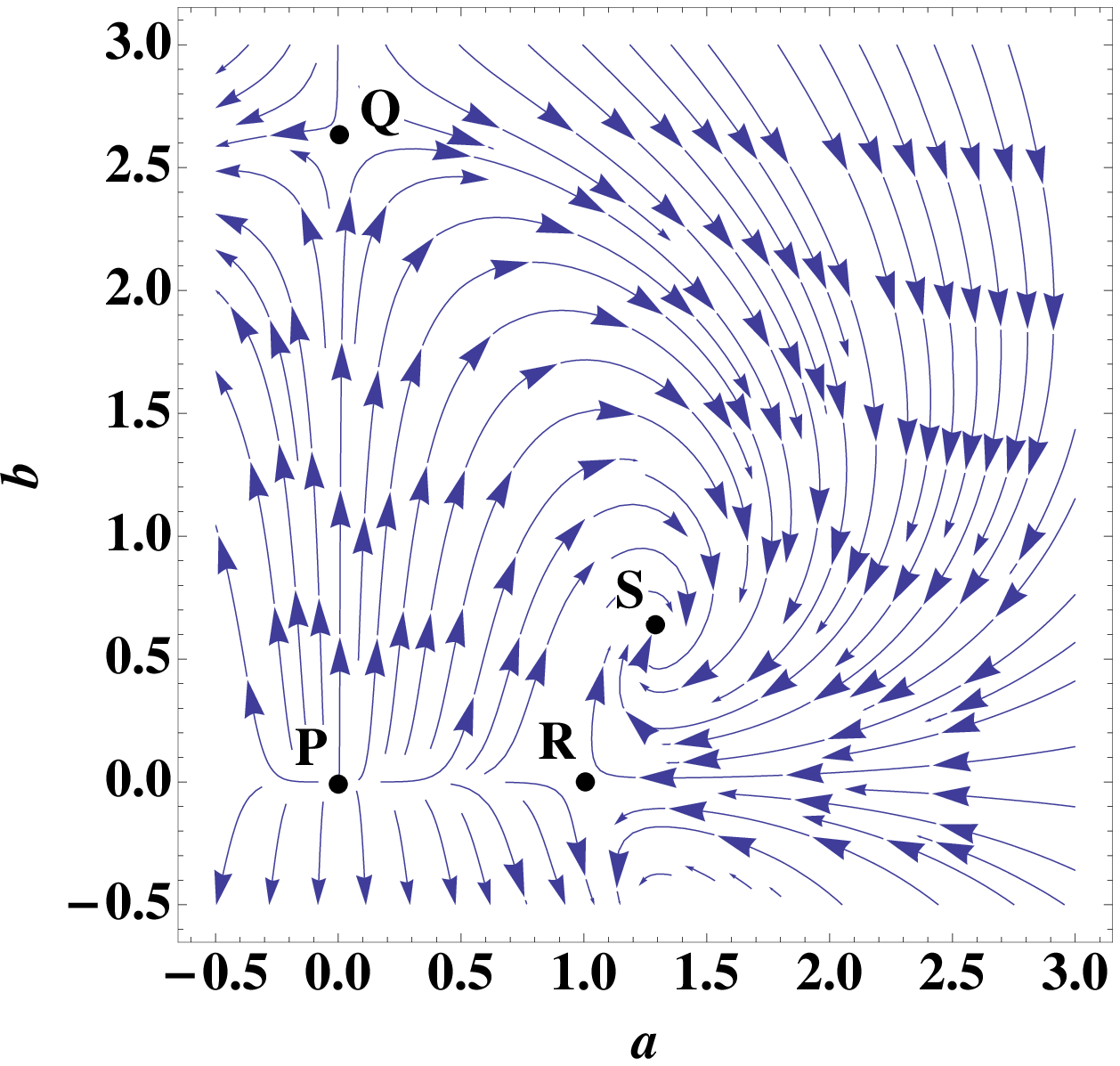}
\caption{Phase plots for simplified penvo ($\Omega=2$) in the Van der Pol plane --- $\gamma=0.5$ (\textit{left}) and $\gamma=1.5$ (\textit{right}): The fixed point $S$ is absent for $0<\gamma<1$ when $R$ is the stable fixed point but shows up as the sole stable fixed point in the positive quadrant when $\gamma>1$.}\label{f:Fig.2}
\end{figure}

First let us focus on the interesting case of $\Omega=2$ in details. The flow equations are
\begin{subequations}
\be
&&\dot{a}=\f{\v a}{2}\left[\left(1-\f{\gamma}{2}\right)-\f{a^2}{4}+\left(\f{\gamma}{2}-\f{1}{4}\right)b^2\right]\\
&&\dot{b}=\f{\v b}{2}\left[\left(1+\f{\gamma}{2}\right)-\f{b^2}{4}-\left(\f{\gamma}{2}+\f{1}{4}\right)a^2\right]\,.
\en
\label{abO2}
\end{subequations}
\noindent There are nine fixed points for Eqs. (\ref{abO2}) which are given by: $
(a^2,b^2)=(0,0);\,(0,4+2\gamma);\,(4-2\gamma,0) \textrm { and }(1+1/\gamma,1-1/\gamma)$.
The pair of eigenvalues of the characteristic matrix obtained by linearising Eqs. (\ref{abO2}) about these fixed points respectively are: ${\v}(2\pm\gamma)/4$; ${\v\gamma}(\gamma+1)/2,-{\v}(\gamma+2)/2$; ${\v\gamma}(\gamma-1)/2,{\v}(\gamma-2)/2$ and $-{\v}/4\pm\v{(5-4\gamma^2)^{1/2}}/4$. Particularly, in the first (positive) quadrant of the $a-b$ phase plane, the real fixed points for $|\gamma|<1$ are (cf. Fig. \ref{f:Fig.2}) three points $P$, $Q$ and $R$ given by $(0,0)$, $(0,+(4+2\gamma)^{1/2})$ and $(+(4-2\gamma)^{1/2},0)$ respectively. $P$ is an unstable node, $Q$ is a saddle point and $R$ is the only stable fixed point. The amplitude of oscillation corresponding to $R$ is $(4-2\gamma)^{1/2}$ which we have used in Fig. \ref{f:Fig.1} to fit the numerical data. (In passing, we mention that when $\gamma<0$, instead of $R$, $Q$ becomes the only fixed point in the first quadrant; but the corresponding stable oscillation's amplitude is still $(4-2|\gamma|)^{1/2}$.) However, the fitting breaks down after $|\gamma|>1$. So, what happens at $|\gamma|=1$? Assume $\gamma=1$ (argument for $\gamma=-1$ goes along similar line). At $\gamma=1$, another real fixed point $S$ given by $((1+1/\gamma)^{1/2},(1-1/\gamma)^{1/2})$ emerges coincided with $R$. As $\gamma$ becomes greater than $1$, an inspection of the corresponding eigenvalues tells us that among $P,Q,R$ and $S$, only $S$ is a stable fixed point in the positive quadrant. The emergence of new stable fixed point fixes the amplitude of the stable oscillations at $(a^2+b^2)^{1/2}\, (\textrm{at } S)=(1+1/\gamma+1-1/\gamma)^{1/2}=\sqrt{2}$ for all $\gamma\ge1$.

The case {$\Omega=4$} can be explained similarly. We have in the Van der Pol plane trajectories governed by
\begin{subequations}
\be
&&\dot{a}=\f{\v a}{16}\left[8-(2+3\gamma)b^2+(\gamma-2)a^2\right]\\
&&\dot{b}=\f{\v b}{16}\left[8-(2+3\gamma)a^2+(\gamma-2)b^2\right]\,.
\en
\label{abO4}
\end{subequations}
\noindent There are nine fixed points for Eqs.~(\ref{abO4}) which are given by: $(a^2,b^2)=(0,0);\,(0,8/(2-\gamma));\,(8/(2-\gamma),0)\textrm{ and }(4/(2+\gamma),4/(2+\gamma))$. The pair of eigenvalues of the characteristic matrix obtained by linearising Eqs. (\ref{abO4}) about the above fixed points respectively are: ${\v}/{2},{\v}/{2}; -2\v,{4\gamma\v}/{2(\gamma-2)};-2\v,{4\gamma\v}/{2(\gamma-2)};$ and $-\v,{2\gamma\v}/{2+\gamma}$. With the help of these information, one can show that the amplitude of the periodic oscillations is $2\sqrt{2}/(2-|\gamma|)^{1/2}$ --- a function which fits well with the numerical data presented in Fig. \ref{f:Fig.1}. For $|\gamma|>2$, there is no stable fixed point of Eqs. (\ref{abO4}).

Finally consider the case: $\Omega\ne 2,4$. When $\Omega$ is an integer, the flow equations in the Van der Pol plane are
\be
\dot{a}=\f{\v a}{2}\left(1-\f{a^2}{4}-\f{b^2}{4}\right);\,\dot{b}=\f{\v b}{2}\left(1-\f{a^2}{4}-\f{b^2}{4}\right)\,.
\label{abOo}
\en
The origin $(a,b)=(0,0)$ is a fixed point of Eqs. (\ref{abOo}). The pair of eigenvalues of the characteristic matrix obtained by linearizing Eq. (\ref{abOo}) about the origin are: ${\v}/{2},{\v}/{2}$. Encircling this unstable origin, there exists a stable limit cycle in the Van der Pol plane: $a^2+b^2=4$. This explains the fixed amplitude of $2$ for the limit cycles in the simplified penvo when $\Omega$ takes integer values other than $2$ and $4$. For any non-integral value of $\Omega$, the harmonic balance procedure adopted here conceptually breaks down. However, straightforward application of any standard perturbation technique such as renormalization group method \cite{CGO96,SBCB11}, Lindstedt-Poincar\'e method (used below) etc. tells us that the corresponding limit cycle's radius remains $2$.

Having completely explained the amplitude of the stable oscillations in the simplified penvo, it is straightforward to find out the correction to the natural frequency ($\omega_0=1$) using Lindstedt-Poincar\'e method \cite{JS}. To this end, we assume following perturbation series:
\be
&&x(t)=x_0(t)+\v x_1(t)+\v^2 x_2(t)+\mathcal{O}(\v^3)\,,\\
&&\omega(t)=\omega_0(t)+\v \omega_1(t)+\v^2\omega_2(t)+\mathcal{O}(\v^3)\,.
\en
Usage of these series in Eq. (\ref{penvo}) leads to following solution for $x_0(t)$:
\be
x_0=A\cos(\omega t)+B\sin(\omega t)\,
\en
using which in the next order equation for $x_1(t)$, and subsequently equating the coefficient of $\cos(t)$ and $\sin(t)$ to zero in order to get rid of secular terms, we arrive at two simultaneous conditions on $\omega_1^2$ as a function of $A$ and $B$.
One can easily argue that --- for the periodic oscillations found using harmonic balance method --- $\omega_1=0$ for all $\Omega$. Thus, the frequency of the stable oscillations in simplified penvo, to the first subleading order in $\v$ is
\be
\omega=1+\mathcal{O}(\v^2).
\en

\begin{figure}
\includegraphics[width=4.2cm]{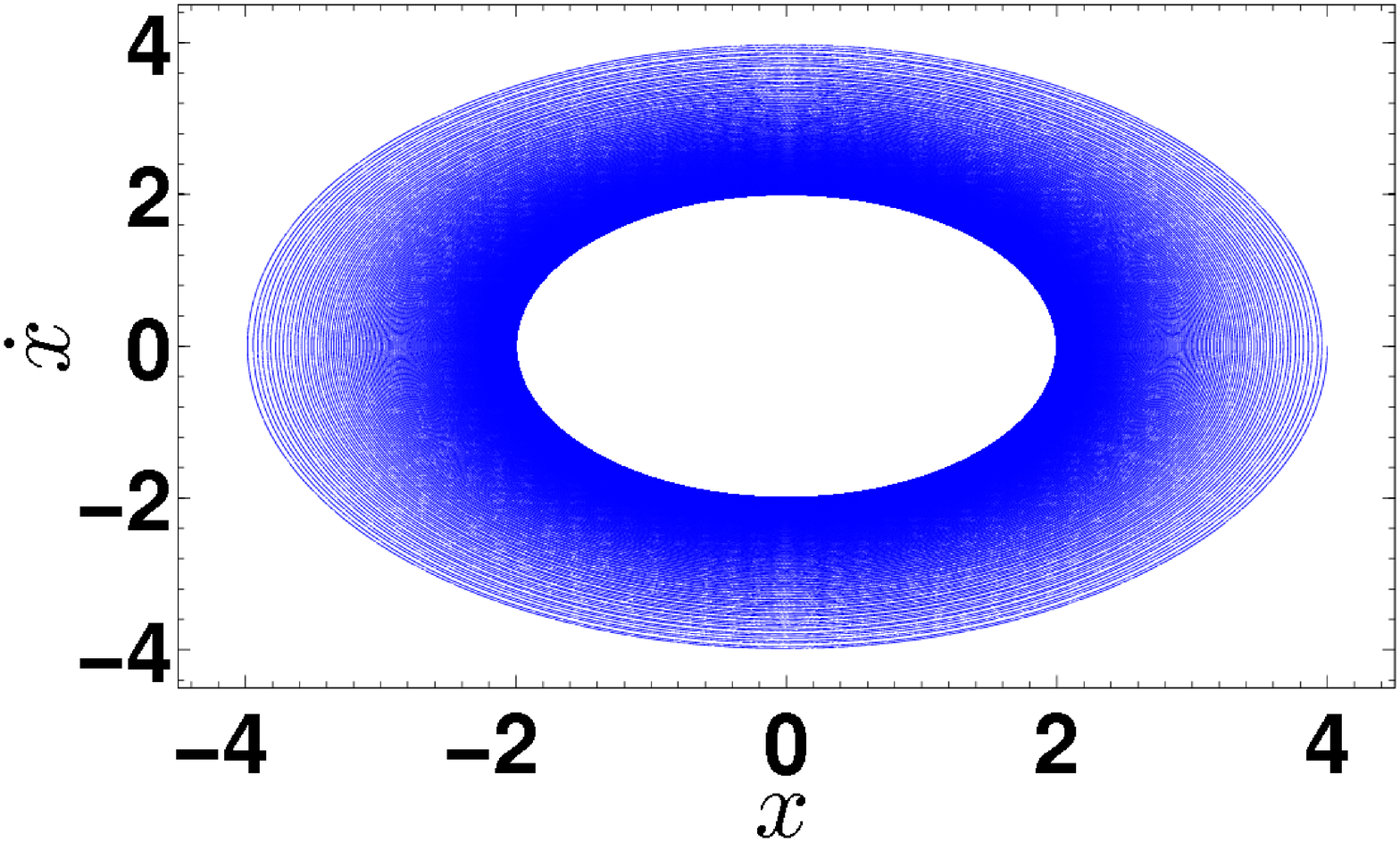}
\includegraphics[width=4.2cm]{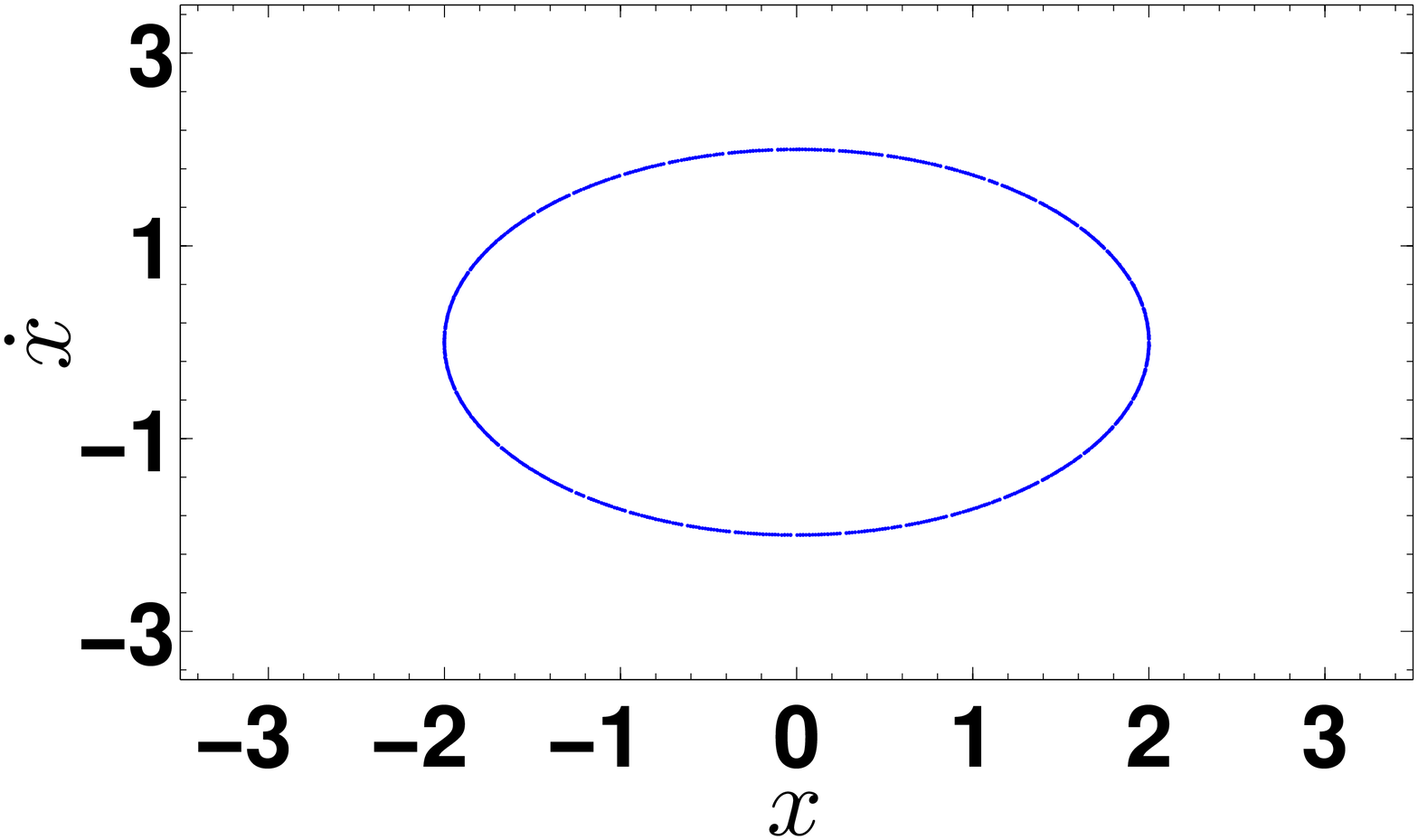}
\caption{Quasiperiodicity in simplified penvo ($\Omega=\sqrt{2}$): (\textit{left}) A trajectory is seen to densely fill up a portion of $x-\dot{x}$ plane of phase space, and (\textit{right}) an elliptical curve traced by the same trajectory in the relevant Poincar\'e section (cf. text).}\label{f:Fig.3}
\end{figure}

With a view to exploring the existence of quasiperiodicity, let us rewrite Eq. (\ref{penvo}) as an autonomous set of three first order ordinary differential equations:
\begin{subequations}
\be
&&\dot{x}_1=x_2\,,\\
&&\dot{x}_2=-x_1-\v(1+\cos x_3)(1-x_1^2)x_2\,,\\
&&\dot{x}_3=\Omega\,.
\en
\end{subequations}
Whether for irrational $\Omega$, the oscillations are quasiperiodic or not, is best addressed by looking at the Poincar\'e section at some plane: $x_3=$constant. Since the third phase space coordinate $x_3$ is the phase of the parametric excitation, the appropriate Poincar\'e section corresponds to looking at the oscillations after every period $2\pi/\Omega$. For periodic oscillation, depending on the ratio between angular frequency of the oscillation and $\Omega$, the Poincar\'e section will consist of finite number of discrete points, whereas any quasiperiodic dynamics will generate infinite number of points on the Poincare section so that a dense curve will be seen --- this exactly is what we see in simplified penvo with irrational $\Omega$ (cf. Fig. \ref{f:Fig.3}).

%

%
%

Let us come back to Eq. (\ref{penvofull}) with $\omega_0=1$ without any loss of generality. First consider the case of $p(t)$ being an odd function so that the Fourier series (\ref{fourier}) of p(t) is an exclusively sine series. One can easily show using harmonic balance --- as used before --- that for such a form of $p(t)$, there doesn't exist a single stable node or spiral in the Van der Pol plane. Consequently, {the limit cycle in Van der Pol equation is destroyed and there is no \textit{limiting} periodic oscillation}. 
When $p(t)$ is a even function, all the sine terms in the Fourier series of $p(t)$ vanish. Rescaling $\v$ and defining $\gamma_{ci}\equiv2a_i/a_0$, we find that, to the leading order, only $\gamma_{c2}$ and $\gamma_{c4}$ can determine the conditions for stable periodic oscillations. Other $\gamma_{ci}$'s are irrelevant for our purpose: when $\gamma_{c2}=\gamma_{c4}=0$, penvo will always yield a limit cycle solution --- $a^2+b^2=4$ --- irrespective of what values other $\gamma_{ci}$'s take.

In the most general case, a complete description of penvo in the oscillating state effectively requires any arbitrary $2\pi$ periodic $p(t)$ to be represented by the following surprisingly small fraction of the full Fourier series (\ref{fourier}):
\be
&&\v p(t)=\v\left(1+\gamma_{c2}\cos 2t+\gamma_{c4}\cos 4t+\gamma_{s2}\sin 2t+\gamma_{s4}\sin 4t\right.\nonumber\\
&&\phantom{p(t)=}\left.+\textrm{irrelevant terms}\right)\,,\label{fourierTr}
\en
where $a_0/2$ has been absorbed in $\v$ and $\gamma_{si}\equiv 2b_i/a_0$. When only `$1$+irrelevant terms' in Eq. (\ref{fourierTr}) survive, the usual Van der Pol oscillator's limit cycle is recovered. While dealing with all four $\gamma$'s simultaneously is cumbersome, we can understand relative importance of the sinusoidal terms in expression (\ref{fourierTr}) by setting any two of the four $\gamma$'s to zero at a time. Out of the six cases that can thus arise, only the following forms of $p(t)$ require independent study: (i) $1+\gamma_{c2}\cos(2t)+\gamma_{c4}\cos(4t)$ and (ii) $1+\gamma_{s2}\sin(2t)+\gamma_{s4}\sin(4t)$. Remaining four relevant combinations of sine and cosine can be reduced to these two forms by suitable rescaling of time $t$ and the concerned $\gamma$'s. Since we do not find any qualitatively new result other than already presented eariler, we merely mention that though the presence of four parameters in $p(t)$ make algebra tedious, one can in principle go through it and theoretically explain the numerical results in a straightforward fashion along a similar line of argument adopted herein for the simpler cases.
%
%

Now we present an idea how a two-state-switch may be constructed using a coupled system of two simplified penvos.
\begin{figure}[h]
\includegraphics[width=6cm]{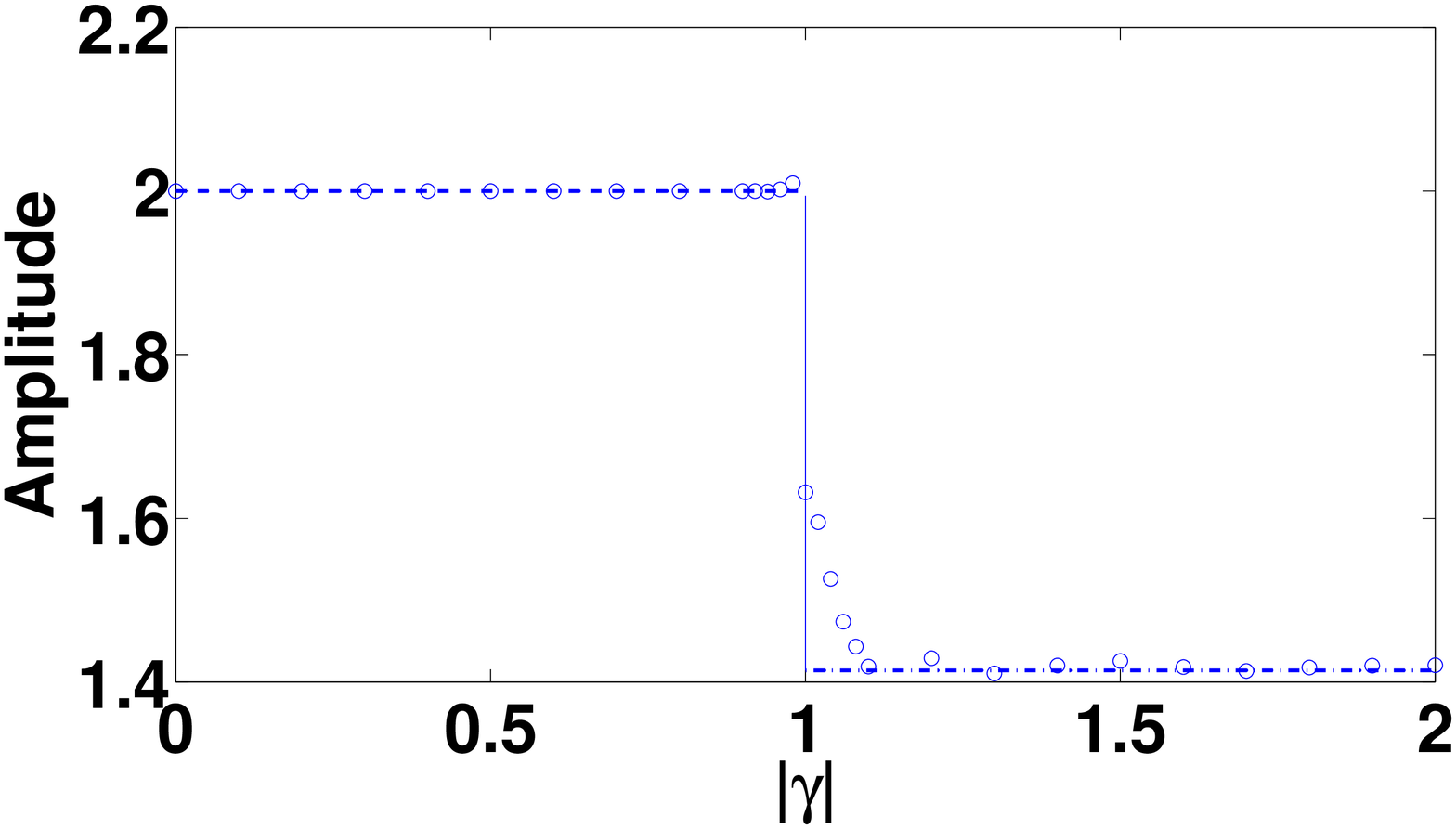}
\caption{Two-state switch in coupled simplified penvos. The behaviour of switch for $|\gamma|\ge1$ is very sensitive to  fluctuations in the value of  `$2\sqrt{2}/A_{y\infty}$'. To compensate this effect, $A_{y\infty}$ has been calculated by taking average of 100 points over the limiting cycles of corresponding oscillations. The dashed line is for amplitude$=2$, the dot-dashed line is for amplitude$=\sqrt{2}$ and circles are the numerical data points.}\label{f:Fig.5}
\end{figure}
Consider,
\begin{subequations}
\be
&&\ddot{x}+\v\left[1+\gamma\cos\left( \f{2\sqrt{2}}{\sqrt{y^2+\dot{y}^2}}t\right)\right](x^2-1)\dot{x}+x=0\,,\qquad\label{penvo1}\\
&&\ddot{y}+\v[1+\gamma\cos(2 t)](y^2-1)\dot{y}+y=0\,.\label{penvo2}
\en
\end{subequations}
When $0\le|\gamma|<1$, Eq. (\ref{penvo2}) results in an oscillation of amplitude $A_{y_\infty}\equiv (y^2_\infty+\dot{y}_\infty^2)^{1/2}=(4-2|\gamma|)^{1/2}$ such that $A_{y_\infty}\in[2,\sqrt{2})$, where subscript $\infty$ refers to the quantities calculated at $t\rightarrow\infty$. This in turn will mean that the angular frequency in the cosine term in Eq. (\ref{penvo1}) lies between $\sqrt{2}$ to $2$ (but not $2$ yet). Hence the amplitude of the stable oscillation one obtains from Eq. (\ref{penvo1}), for $\gamma\in[0,1)$, is always $2$. When $|\gamma|\ge1$, Eq. (\ref{penvo2}) yields a periodic (or quasiperiodic) oscillation of amplitude $\sqrt{2}$ and hence the angular frequency in the cosine term in Eq. (\ref{penvo1}) is fixed at $2$. Consequently, the amplitude of the stable oscillations one obtains from Eq. (\ref{penvo1}) is fixed at $\sqrt{2}$ for $|\gamma|\ge1$. In short, the amplitude of the stable oscillations generated by Eq. (\ref{penvo1}) \textit{switches} from $2$ (for $|\gamma|<1$) to $\sqrt{2}$ (for $|\gamma|\ge1$) at $|\gamma|=1$ (cf. Fig. \ref{f:Fig.5}).

To conclude, we emphasize that in view of this study on Van der Pol oscillator, it is of immediate interest to revisit different Li\'enard systems displaying limit cycle oscillations and investigate the effect of parametric excitation on the non-linear terms. We suspect that results obtained here could be generic and, the resonance and the anti-resonance could be encountered in other such systems as well. Such resonance and anti-resonance can have direct implications on every system across disciplines that uses one or the other form of such oscillators for its mathematical modeling.

\begin{acknowledgements}
We thank Jayanta K. Bhattacharjee, Dhrubaditya Mitra and Shantanu Mukherjee for useful discussions. SC acknowledges financial support from the Danish Research Council through FNU Grant No. 505100-50 - 30,168.
\end{acknowledgements}
%

%
\end{document}